# Coagulation model for ion channels in the lipid membranes

I.A. Larionov *, A.L. Larionov *, E.E. Nikolsky [†‡], M.N. Shneider [§]

*Institute of Physics, Kazan Federal University, Kremlevskaya str. 18, Kazan 420008, Russia
[†]Kazan Institute of Biochemistry and Biophysics, Russian Academy of Sciences, Kazan 420008, Russia
[‡]Kazan State Medical University, Butlerov str. 49, Kazan, 420012 Russia
[§]MAE Department, Princeton University, Princeton, NJ 08544, USA

**ABSTRACT** A two-dimensional (2D) model is developed to describe the growth of spotty patterns (domains) in the lipid membranes which result from the coagulation of ion channels. It is assumed that the ion channels can coagulate when activated, e.g. the neurotransmitter may act as the surface-active substance for some ion channels in the lipid matrix or the activation may result in increase of polarity of the ion channels that give rise to their coagulation process due to the Brownian diffusive motion. The results show that the domain radius $r_D$ of the ion channels scales with time as $r_D \sim time^{0.4}$. The slowing and subsequent saturation of the domain radius at longer time periods are in qualitative agreement with experiments.

It is known that transmembrane protein channels are not fixed in the membrane, but are subject to free lateral diffusion [1-5]. In the postsynaptic membrane, one quantum of the neurotransmitter activates about 1000 ion channels [6,7], and they can assemble into stable domains (clusters) with a number of channels in the domain up to several hundreds. The characteristic time for the domains formation and their decay is of the order of a few seconds. The domain formation is studied in experiments, see, e.g., [8], however, the qualitative and quantitative theory of this interesting and important phenomenon is still in the focus of discussion [8-14]. In the present paper, we show that the domain formation process can be a manifestation of Brownian coagulation and give quantitative characteristics of the growth process of domains for typical active lipid membranes.

We assume that the activation of ion channels acquire the ability to form bonds and stick together as a result of collision. For ligand-activated channels, this can be caused, for example, by the action of the neurotransmitter, including as a surfactant. Activation of the ion channels may also result in increase of their polarity [14]. This leads to interaction (attraction) of the ion channels with force that is known to decrease rapidly with distance.

Randomly "collided" channels are stuck together. The resulting domain has a large lateral surface and, hence, lower mobility. Therefore, spotty patterns of transmembrane channels formed accidentally during the initial stages of the membrane activity can be considered as nearly static initial nuclei and growth centers of the gigantic spotty clusters (domains) observed in experiments. A characteristic estimate of the formation time of the embryonic domain as a result of the Brownian motion of the transmembrane channels, which are subject to free lateral diffusion, gives

$$\tau_i \sim l_0^2/D = 10^{-3} \text{ s}.$$

Here $l_0 \sim 10$nm is the average initial distance (spacing) between the outer surfaces of the ion channels, which have an outer radius of $R_1 = 4$ nm [15] and a typical value of the lateral diffusion coefficient is $D = 10^{-13}$ m$^2$/s [1-5]. We will assume that single activated ion channels come to our arbitrarily chosen "clustering center" from a region with a radius equal to or less than $0.5R_2$ [6,16,17]. This bound eliminates the possibility of activated ion channels diffusing across the circle with radius $R_2 \approx 1$ μm in the time interval considered. The $R_2$ boundary is set by the knowledge that one quantum of the neurotransmitter activates ion channels in a region on the order of 1 μm$^2$ (see, e.g., [6,16,17] and Fig. 1).

The process of rapid domain size growth will end after a time interval of the order of

$$\tau_D \sim R_2^2 / 4D \approx 2.5 \text{ s}.$$

As no new channels are generated, the number and density of free channels in the membrane is reduced and the domain growth slows down and subsequently, ceases.

The problem of the clustering of freely diffusing transmembrane channels reduces to the Brownian coagulation equation described by Smoluchowski's diffusion equation [18]. In this letter we will adopt for ion channels a coagulation model in 2D geometry. The 2D diffusion equation for the ion channels density $n(x, y, t)$ is

$$\frac{\partial n(x,y,t)}{\partial t} = D\left(\frac{\partial^2 n(x,y,t)}{\partial x^2} + \frac{\partial^2 n(x,y,t)}{\partial y^2}\right), \quad (1)$$

Or, for isotropic case in cylindrical coordinates

$$\frac{\partial n(r,t)}{\partial t} = \frac{D}{r}\frac{\partial}{\partial r}\left(r\frac{\partial n(r,t)}{\partial r}\right), \quad (2)$$

where in Cartesian coordinates $r^2 = x^2 + y^2$.

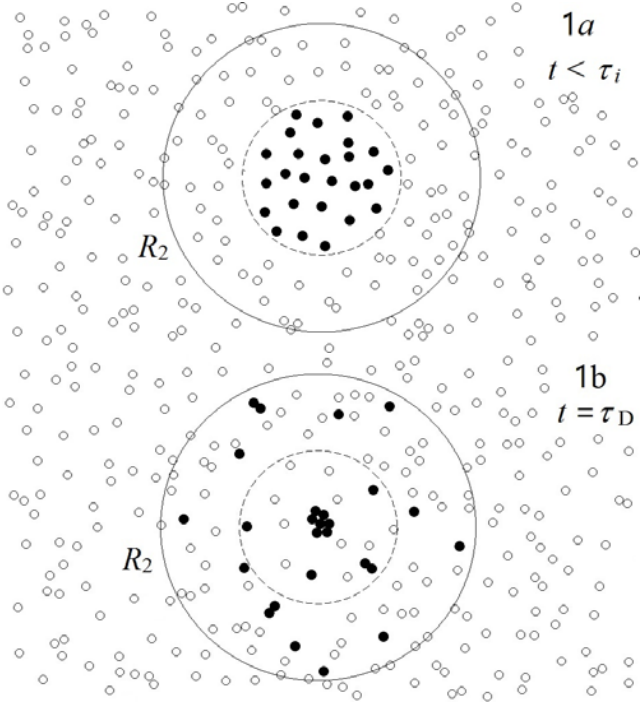

FIGURE 1. Sketch of the ion channels layout at the initial (*a*) and the final moment (b) of the domain growth process. Filled and open circles show the activated and inactivated ion channels, respectively. The inactivated ion channels (open circles) are beyond our consideration. The actual number of activated ion channels in the calculations is $0.25\pi n_0 R_2^2 \approx 1570$.

For the 2D (or the 1D in cylindrical coordinates) case, in analogy with initial and boundary conditions for 3D case, we use the solution [19] for the density of particles in the domain. With the initial condition $n(r, t = 0) = f(r)$, and the boundary conditions for $t > 0$, $n(r = R_1, t) = 0$, and $n(r = R_2, t) = g_2(t)$, the solution has the form [19]:

$$n(r,t) = \int_{R_1}^{R_2} f(\xi) G(r,\xi,t) d\xi - D \int_0^t g_2(\tau) \Lambda_2(r, t-\tau) d\tau, \quad (3)$$

where

$$G(r,\xi,t) = \frac{\pi^2}{2R_1^2} \sum_{k=1}^{\infty} \frac{\mu_k^2 J_0^2(s\mu_k) \Psi_k(r) \Psi_k(\xi) \xi}{J_0^2(\mu_k) - J_0^2(s\mu_k)} \exp\left(-\frac{D\mu_k^2 t}{R_1^2}\right),$$

with

$$\Psi_k(r) = Y_0(\mu_k) J_0\left(\frac{\mu_k r}{R_1}\right) - J_0(\mu_k) Y_0\left(\frac{\mu_k r}{R_1}\right),$$

where $s = R_2/R_1$, $J_0(\mu)$ and $Y_0(\mu)$ are the Bessel functions, $\mu_k$ - are the positive solutions of the equation:

$$J_0(\mu) Y_0(s\mu) - J_0(s\mu) Y_0(\mu) = 0,$$

and

$$\Lambda_2(r,t) = \frac{\partial}{\partial \xi} G(r,\xi,t)\Big|_{\xi=R_2}.$$

Here the uniform sparse initial ion channel density $f(\xi) = n_0 = const$, and the second boundary condition function $g_2(\tau)$ may depend on the model under study $0 \leq g_2(\tau) \leq n_0$. In general $g_2(\tau)$ may be a decreasing function of time $\tau$ in the form $g_2(\tau) = n_0 exp(-\kappa\tau)$ since there are no newly created ion channels in the vicinity of the coagulation center. For simplicity we will consider only the simplest case with $g_2(\tau) = 0$, i.e. the contribution from the first term of Eq. (3). This means that no activated ion channels cross the cirle with radius $R_2$ within the time periods considered.

The ion channels flux on the domain perimeter is [18]

$$j(t) = D\left(\frac{\partial n(r,t)}{\partial r}\right)_{r=r_D}, \quad (4)$$

where $r_D$ is the domain (coagulation) radius. The total number of ion channels crossing the domain perimeter $r = r_D = R_1\sqrt{N}$ per time, i.e. the rate of the coagulation events is then given by

$$M(t) = \oint j(t) \, dL = j(t) \cdot 2\pi R_1 \sqrt{N}$$

and the number of ion channels in the domain is

$$N = \int_0^{time} M(t) \, dt = 2\pi R_1 \sqrt{N} \int_0^{time} j(t) \, dt. \quad (5)$$

We have to take into account the relative independent diffusive motion of the domain and single ion channels where the domain diffusion coefficient scales with its radius as $D_D \propto r_D^{-1}$ (see, e.g., [2,8]) at least in the large $r_D$ limit. Therefore we replace $D$ with $D_{eff}$, where $D_{eff}$ is given by

$$D_{eff} = D + D_D = D\left(1 + \frac{1}{\sqrt{N}}\right)$$

This gives us the equation for the number of ion channels in the domain

$$\sqrt{N} = 2\pi R_1 D\left(1 + \frac{1}{\sqrt{N}}\right) \int_0^{time} \left(\frac{\partial n(r,t)}{\partial r}\right)_{r=r_D} dt, \quad (6)$$

and the domain (coagulation) radius $r_D = R_1\sqrt{N}$. The numerical value of $R_1$ determines the maximum density of ion channels in the domain. For $R_1 = 4$ nm [15] it corresponds to $\approx 18\,000$ ion channels/$\mu m^2$.

We solved Eq. (6) by the least squares method and the results of the calculations are shown in Fig. 2 with the following parameters set: $n_0 = 2\,000/\mu m^2$ [17], $R_1 = 4$ nm, $R_2 = 1000$ nm, and $g_2(\tau) = 0$. The values of the lateral diffusion coefficients are $D = 10^{-12}\,m^2/s$, $D = 10^{-13}\,m^2/s$, and $D = 10^{-14}\,m^2/s$, from left to right, respectively. The calculations show that the number of ion channels in the central domain $N$ and the domain radius $r_D$ scale with time and the diffusion coefficient $D$ as

$$N^{0.5} = \frac{r_D}{R_1}; \quad 2.6 \times 10^6 \times (D \times time)^{0.4}.$$

The relative independent diffusive motion of the domain and single ion channels affects the domain radius growth factor only a little except at the initial moments, that are beyond our interest. The number of ion channels in the domain is finally $N≈520$; this means that around 1/3 of the single ion channels, according to our estimate, form the central large domain. The ion channel layout in the domain formed by the coagulation processes due to the Brownian diffusive motion is in the disordered form. The increase of polarity of the activated ion channels may also affect their layout in the domain as reported in [14]. Investigation of the polarity or charge distribution within the activated ion channels are higly desirable both experimentally and theoretically.

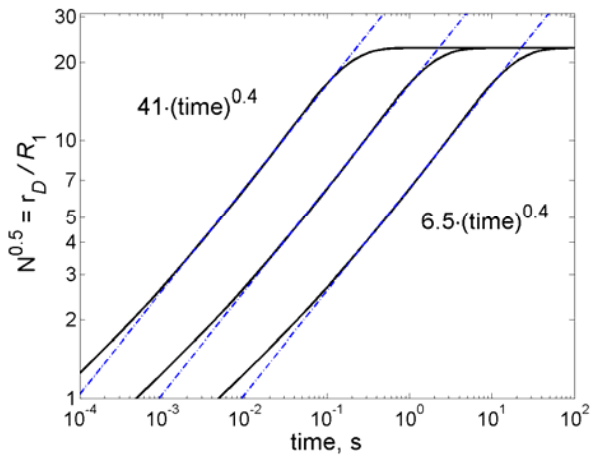

FIGURE 2. The domain radius $r_D$ (or $N^{0.5}$, where $N$ is the total number of ion channels in the domain) *versus* time (solid lines). The values of the lateral diffusion coefficients are $D=10^{-12}$ m$^2$/s, $D=10^{-13}$ m$^2$/s, and $D=10^{-14}$ m$^2$/s, from left to right, respectively. The dash-dotted lines show $N^{0.5}=r_D/R_1=2.6×10^6×(D×time)^{0.4}$.

In this Letter we showed that giant domain-spots, consisting of transmembrane protein channels observed in experiments with active membranes can form as a result of Brownian coagulation. In this case, the characteristic domain formation time is of the order of fractions of a second to a few seconds, depending on the value of the lateral diffusion coefficient. This is in agreement with known experimental data. It should be expected that at the end of the active phase of excitation (when, in our opinion, the binding mechanism ceases to function), the diffusion effect leads to smearing the domain and relaxation to a uniform distribution of the transmembrane channels occurs approximately within the same times.

**AUTHOR CONTRIBUTIONS**
M.N.S. proposed the model, I.A.L. and A.L.L. performed the calculations, E.E.N. initiated and supervised the work. All authors jointly wrote the article.